\documentclass[aps,prl,amssymb,showpacs,twocolumn]{revtex4}
\usepackage{graphicx}

\begin{document}

\preprint{IMSc/2003/10/34}
\preprint{AEI--2003--082}
\preprint{CGPG--03/10--4}
\preprint{gr--qc/0311003}

\title{Quantum suppression of the generic chaotic behavior close to 
cosmological singularities}

\author{Martin Bojowald}
\email{mabo@aei.mpg.de}
\affiliation{Max-Planck-Institut f\"ur Gravitationsphysik, Albert-Einstein-Institut, Am M\"uhlenberg 1, D-14476 Golm, Germany}
\affiliation{Center for Gravitational Physics and Geometry, The Pennsylvania
State University, 104 Davey Lab, University Park, PA 16802, USA}

\author{Ghanashyam Date}
\email{shyam@imsc.res.in}
\affiliation{The Institute of Mathematical Sciences,
CIT Campus, Chennai-600 113, INDIA}


\newcommand{\lP}{\ell_{\mathrm P}}

\newcommand{\md}{{\mathrm{d}}}
\newcommand{\Kern}{\mathop{\mathrm{ker}}}
\newcommand{\tr}{\mathop{\mathrm{tr}}}
\newcommand{\sgn}{\mathop{\mathrm{sgn}}}

\newcommand*{\R}{{\mathbb R}}
\newcommand*{\N}{{\mathbb N}}
\newcommand*{\Z}{{\mathbb Z}}
\newcommand*{\Q}{{\mathbb Q}}
\newcommand*{\C}{{\mathbb C}}

\begin{abstract}
In classical general relativity, the generic approach to the initial
singularity is very complicated as exemplified by the chaos of the
Bianchi IX model which displays the generic local evolution close
to a singularity. Quantum gravity effects can potentially change the
behavior and lead to a simpler initial state. This is verified here in
the context of loop quantum gravity, using methods of loop quantum
cosmology: the chaotic behavior stops once quantum effects become
important. This is consistent with the discrete structure of space
predicted by loop quantum gravity.
\end{abstract}

\maketitle

According to the celebrated singularity theorems of classical general
relativity, the backward evolution of an expanding universe leads to a
singular state where the classical theory ceases to apply. An
extensive analysis of the approach to the singularity, in the general
context of inhomogeneous cosmologies, has culminated in the
BKL-scenario \cite{BKL}.  According to this scenario, as the
singularity is approached, the spatial geometry can be viewed as a
collection of small patches each of which evolves essentially
independently as a homogeneous model, most generally the Bianchi IX
model. This is justified by the observation that interactions between
the patches are negligible because time derivatives dominate over
space derivatives close to a singularity.

\begin{figure}[b]
\includegraphics[width=8cm,height=6cm]{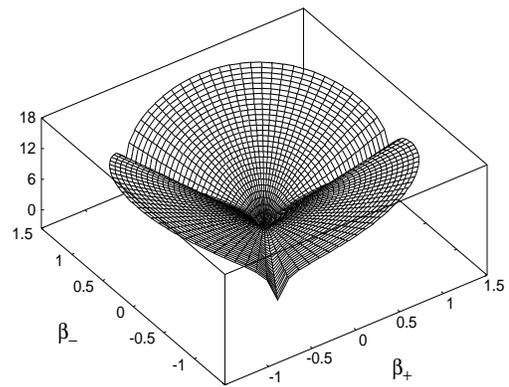}
\caption{The potential is shown for $V= 30$ in Planck units.
Along the $z$-axis is plotted the logarithm of the potential (shifted 
so that it is larger than 1 everywhere).  Its form does not change
with volume due to the factorized volume dependence.}
\label{ClassFig}
\end{figure}

The approach to the singularity of a Bianchi IX model is described by
a particle moving in a potential with exponential walls (corresponding
to the increasing curvature) bounding a triangle
(Fig.~\ref{ClassFig}). During its
evolution the particle is reflected at the walls resulting in an
infinite number of oscillations (of the scale factors) when the
singularity is approached. This classical behavior can be shown to
lead to a chaotic evolution by using an analogy with a billiard valid
in the asymptotic limit close to the singularity \cite{chaos}.

To appreciate implications of the chaotic approach to the Bianchi IX
singularity, observe that at any given time the spatial slice can be
decomposed into a collection of almost homogeneous patches, the size
of the patches being controlled by the magnitude of space derivatives
in the equations of motion. During subsequent evolution when the
curvatures grow, these patches have to be subdivided to maintain the
homogeneous approximation.  This subdivision is also controlled by
evolution of the individual patches.  Since the patches are
homogeneous only to a certain approximation, a subdivision of a given
patch at a certain time leads to new patches with slightly different
initial conditions. The chaotic approach to the Bianchi IX singularity
then implies that their geometries will depart rapidly from each
other, and the patches have to be fragmented more and more the closer
one comes to the singularity. This rapid fragmentation suggests a very
complicated and presumably fractal structure of the spatial geometry
at the classical singularity.  Note that both of these features, the
breaking up of a spatial slice into approximately homogeneous patches
and the unending oscillatory approach to the singularity of individual
patches, are ultimately consequences of the unbounded growth of the
spatial curvature, i.e. the singularity.

However, the classical evolution towards the singularity is expected
to break down when the curvatures become too large and to be replaced
by quantum dynamics. If we truncate the classical model at a certain
lowest volume, technically chaos does not occur. It is a natural
question to ask if the qualitative features of the chaotic approach
continue to survive quantum modifications. The answer to this
question is neither obvious nor independent of concrete theories of
quantum gravity and must be explored within the context of a specific
candidate theory.

For instance, effective actions with extra terms and additional fields
motivated from string or M theories do not lead to a non-chaotic
behavior \cite{StringChaos} and thus retain their implications for
fragmentation in the context of a general inhomogeneous singularity.
On the other hand, loop quantum gravity \cite{Rov:Loops,ThomasRev}
predicts a discrete structure of space. For these theories, an
unlimited fragmentation of space would be inconsistent with the
discrete structure and one would expect a qualitative modification of
the chaotic approach.  This translates into a self consistency test
for such theories. {\em Thus, a quantum theory of gravity with a
discrete structure of space must, for self-consistency, provide a
mechanism which prevents the unlimited fragmentation in the expected
general approach to a classical singularity.}

In this paper we use methods of loop quantum cosmology
\cite{LoopCosRev,HomCosmo,Spin}, a part of loop quantum gravity, to study 
this issue. This allows us to obtain explicit, non-perturbative
modifications of the classical behavior at small volume which can be
analyzed for their implications for chaos. Loop quantum cosmology has
already lead to a resolution of conceptual problems such as a
non-singular evolution \cite{Sing} and also given a new scenario for
inflation \cite{Inflation} which is based on the small-volume
modifications. The proof of absence of singularities extends to
homogeneous models, in particular the Bianchi IX model
\cite{HomCosmo,Spin}.  The investigations of this paper will provide a
new consistency check of loop effects and their physical viability.
Specifically, quantum modifications for the vacuum Bianchi IX model
(central to the issue of a chaotic approach) are presented and shown
to prevent the chaotic behavior.

The dynamics of the Bianchi IX model can be formulated on its
minisuperspace spanned by the positive scale factors $a_I$,
$I=1,\ldots, 3$, related to the diagonal metric components
$g_{II}=a_I^2$. It is given by the Hamiltonian constraint 
\begin{eqnarray}
 H&=& \frac{2}{\kappa}\left[~(\Gamma_J\Gamma_K-\Gamma_I)a_I
   -\frac{1}{4} a_I\dot{a}_J\dot{a}_K + \text{cyclic}
   \right]\, \label{HH}
\end{eqnarray}
where $\kappa=8\pi G$ is the gravitational constant and
\begin{equation} \label{SpinConn}
 \Gamma_I = \frac{1}{2}\left(\frac{a_J}{a_K}+ \frac{a_K}{a_J}-
   \frac{a_I^2}{a_Ja_K}\right)
 \mbox{ for $\epsilon_{IJK}=1$}
\end{equation}
are the spin connection components. Thus, the potential term obtained
from (\ref{HH}) is given by,
\begin{equation}\label{ClassPot}
W(a_1, a_2, a_3)  =  \frac{1}{2}\left[ \sum_I a_I^4  -  2
(a_1a_2a_3)^2 \sum_I a_I^{-2} \right] 
\end{equation}

In order to diagonalize the kinetic term of (\ref{HH}) one can
introduce the Misner variables \cite{Mixmaster}: the logarithmic
volume $\Omega := -\frac{1}{3} \log V = -\frac{1}{3}\log(a_1a_2a_3)$
and the anisotropies $\beta_{\pm}$, introduced via $ a_1 :=
e^{-\Omega+\beta_++\sqrt{3}\beta_-} , a_2 :=
e^{-\Omega+\beta_+-\sqrt{3}\beta_-} , a_3 := e^{-\Omega-2\beta_+} \,.$
In terms of these variables the potential (\ref{ClassPot}) is obtained
as,
\begin{eqnarray} \label{PotFar}
{\cal W}(\Omega, \beta_+, \beta_-) & = & \frac{1}{2}e^{-4\Omega} \left[
e^{-8\beta_+} - \right. \nonumber \\
& & 4 e^{- 2\beta_+}
\text{cosh}( 2 \sqrt{3} \beta_- ) + \nonumber \\
& & \left. 2 e^{4\beta_+}\left\{\text{cosh}(4 \sqrt{3} \beta_-) - 1
\right\}
\right] 
\end{eqnarray}
Evidently, the $\Omega$-dependence factorizes and one obtains an
anisotropy potential ${\cal V}(\beta_+,\beta_-)$ which exhibits
exponential walls for large anisotropies as displayed in
Fig.~\ref{ClassFig}.  A typical wall can be derived from the potential
by setting $\beta_-=0$ and taking $\beta_+$ to be negative, e.g.
$W(a_1,a_1,V/a_1^2)=\frac{1}{2}e^{-4\Omega}(e^{-8\beta_+}-
4e^{-2\beta_+})\sim \frac{1}{2}e^{-4\Omega-8\beta_+}$.

The wall picture implies that the universe, described by a particle
moving in such a potential, runs through almost free (Kasner) epochs
where the potential can be ignored, interrupted by reflections at the
walls where the expansion/contraction behavior of different directions
changes. The infinite number of these reflections implies that the
system behaves chaotically.

One can see that the infinite height of the walls is a consequence of
the diverging intrinsic curvature in $\Gamma_I$ for small and large
$a_I$. In the classical evolution each $a_I$ will eventually become
arbitrarily small as the singularity is approached, e.g.\ $a_1$ and
$a_2$ for the typical wall above. However, at a certain stage of the
evolution quantum gravity is expected to become important which will
lead to a modification of the behavior. Clearly, a necessary condition 
for the chaotic behaviour to be prevented by the quantum modification is 
that quantum gravity should effectively contain an upper limit on the 
curvature such that the walls would have only finite height, changing the 
whole scenario.

A quantum theory where such a maximal curvature {\em follows} 
is loop quantum cosmology \cite{InvScale,QSDV}. The origin of
this upper bound for the curvature lies in the quantization of the
relevant quantities and can be compared conceptually to the finite
ground state energy of the hydrogen atom obtained after quantization.
In the classical equations of motion one can incorporate this
feature of loop quantum cosmology by replacing the $\Gamma_I$ with
effective coefficients which are derived from the quantization
\cite{Spin}.  This leads to the corresponding effective potential.

The new, effective potential is more complicated than the original
one.  Nonetheless, for large volume, the effective potential
approximates the original one. It is significantly different for small
volumes and is responsible for the breakdown of chaos. The volume
dependence does not factorize, making the analysis of the classical
motion (with volume as internal time) harder. One should also keep in
mind that the analysis done here uses only the effective classical
description which includes some {\em non-perturbative} quantum effects
in the potential. The full quantum evolution is much more complicated
and is given by a partial difference equation for the wave function
\cite{Spin}.

Using the effective potential in classical equations of motion is
sufficient to shed light on the above consistency requirement, even
though it is not valid close to the classical singularity.  The
underlying dynamical equation of loop quantum cosmology contains a
parameter, $j$, that appears as a quantization ambiguity
\cite{Ambig}. Its value controls the size of the universe where the
maximal curvature is attained. By choosing it to be sufficiently
large, one can move the quantum effects in the effective potential
into the semiclassical domain. Those large values may not be expected
or natural from a physical point of view, but they allow us to study
the consistency issue in a simplified setting. A necessary requirement
for consistency of the quantum theory then is that the effective
classical description stops the Bianchi IX oscillations at some
volume.

To be more explicit, one first has to introduce densitized triad
variables $|p^I|=V/a_I$, with the volume $V=a_1a_2a_3=
\sqrt{|p^1p^2p^3|}$, which become basic operators in loop quantum
gravity. For convenience, we have taken them to be dimensionless
by setting $\frac{1}{2}\gamma\lP^2=1$ in the notation of
\cite{Spin}.  Their inverses $|p^I|^{-1}$ do have well-defined
quantizations as finite operators, despite the classical curvature
divergence at $p^I=0$ which is eliminated by quantum effects
\cite{InvScale}. One can model its effect in the classical behavior by
replacing $(p^I)^{-1}$ in the spin connection components by a function
$F(p^I/2j)$ to get an effective spin connection which includes
non-perturbative quantum effects. The function $F$ is derived
directly from a quantization \cite{Ambig}. The ambiguity parameter $j$
appears explicitly and controls the peak of $F$. 

\begin{figure}[ht]
\begin{center}
 \includegraphics[width=7cm,height=5cm,keepaspectratio]{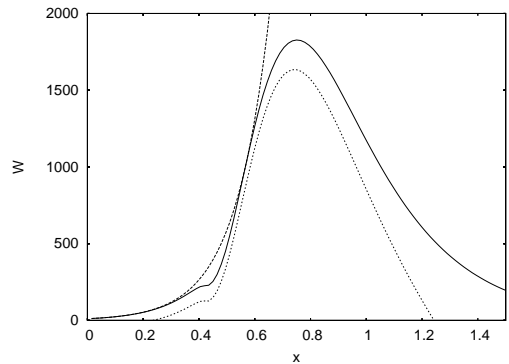}
\end{center}
\caption{Effective wall (\ref{Wall}) of finite height for $j=1$, $V=10$ as a
function of $x=-\beta_+$, compared to the classical exponential wall
(upper dashed curve). Also shown is the exact wall
$W_j(p^1,p^1,(V/p^1)^2)$ (lower dashed curve), which for $x$
smaller than the peak value coincides well with the approximation
(\ref{Wall}) up to a small, nearly constant shift.}
\label{WallProf}
\end{figure}

The potential $W$ then gets replaced by an effective one, $W_j$, by
using the effective instead of the classical spin connection
components. One can again obtain the typical wall behavior, this time
by evaluating $W_j(p^1,p^1,(V/p^1)^2)$ as a function of $p^1$ at fixed
volume.  At these arguments, with $p^1$ large enough but $V/p^1$ not
too small, the complicated expression simplifies to just two dominant
terms.  As a function of $p^1$ and the volume, the typical wall
becomes (see Fig.~\ref{WallProf})
\begin{equation} \label{Wall}
 W_j(p^1,p^1,2jq) \approx 
\frac{V^4F^2(q)}{32j^4q^2} \{3
- 2q F(q)\} ,
\end{equation}
where we have used the notation $q := \frac{1}{2j}(V/p^1)^2$.

For values at the peak or larger, $F(q)\sim q^{-1}$, and the classical
wall $\frac{1}{2}e^{-4\Omega-8\beta_+}$ is reproduced. The peak of the
finite walls is reached for a constant argument $q$ of $F$ which in
Misner variables implies that $e^{-2\Omega+2\beta_+}$ is
constant. Thus, the wall maxima lie on the line $\beta_+=\Omega+{\rm
const}$ in the classical phase space, and the height of the wall drops
off as $e^{-12\Omega}\propto V^4$ with decreasing volume. At very
small volume, however, the walls collapse even more rapidly, and a
numerical analysis shows that the potential becomes negative
everywhere at a dimensionless volume of about $(2.172 j)^{3/2}$ in
Planck units, i.e.\ just around the elementary discrete volume for the
smallest value $j=1/2$.

\begin{figure}
\begin{center}
\includegraphics[height=6cm]{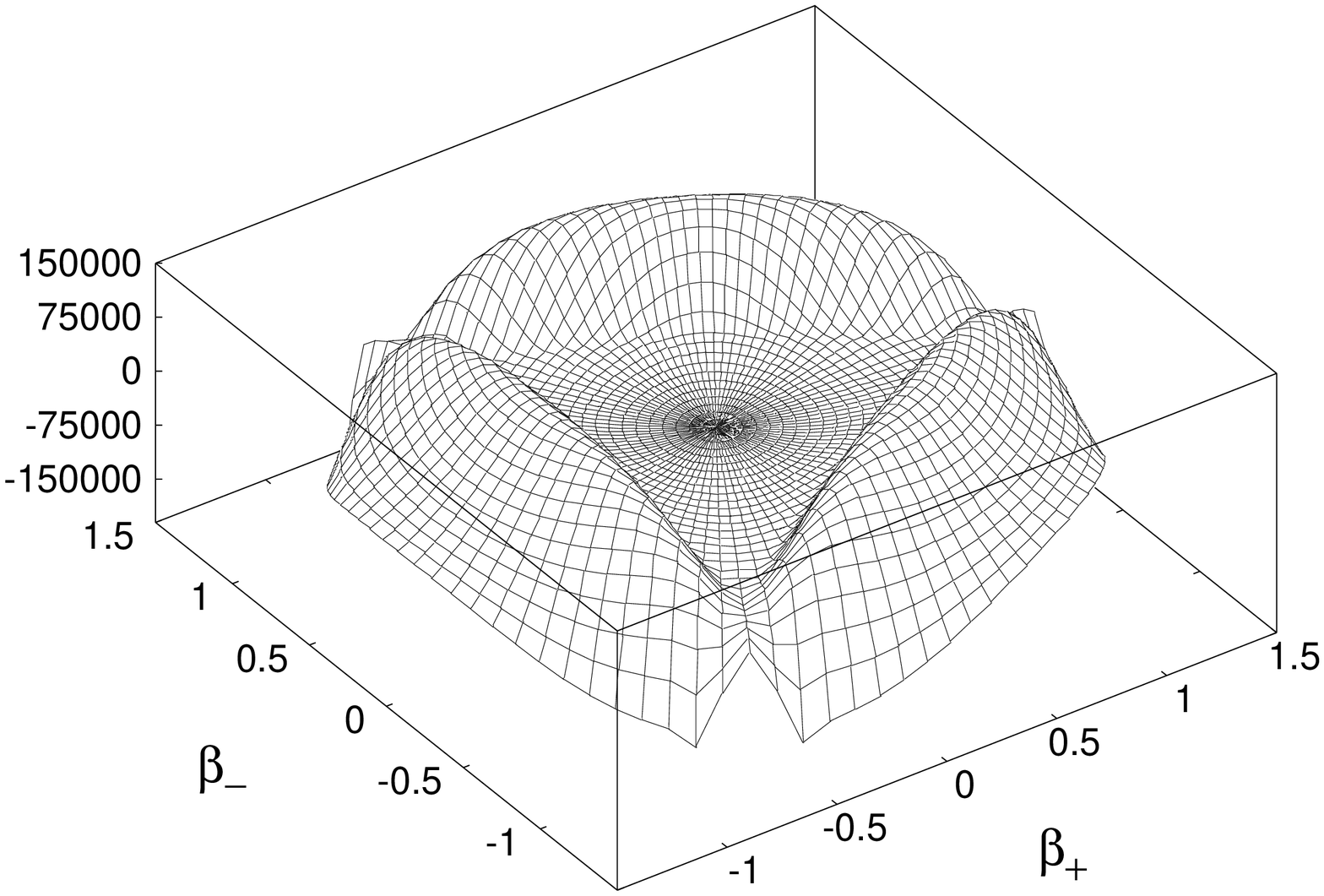}
\includegraphics[width=8cm,height=6cm]{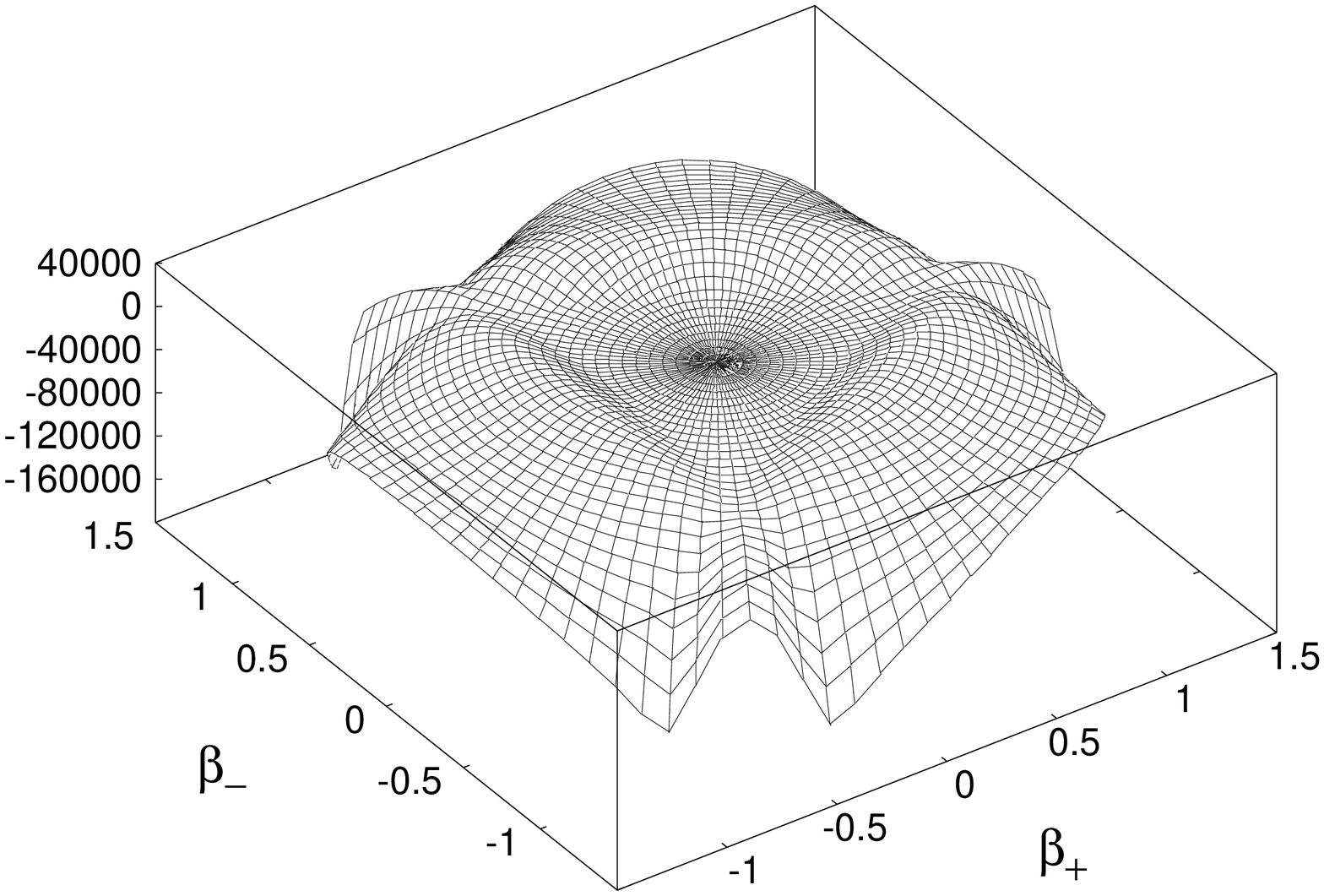}
\includegraphics[width=8cm,height=6cm]{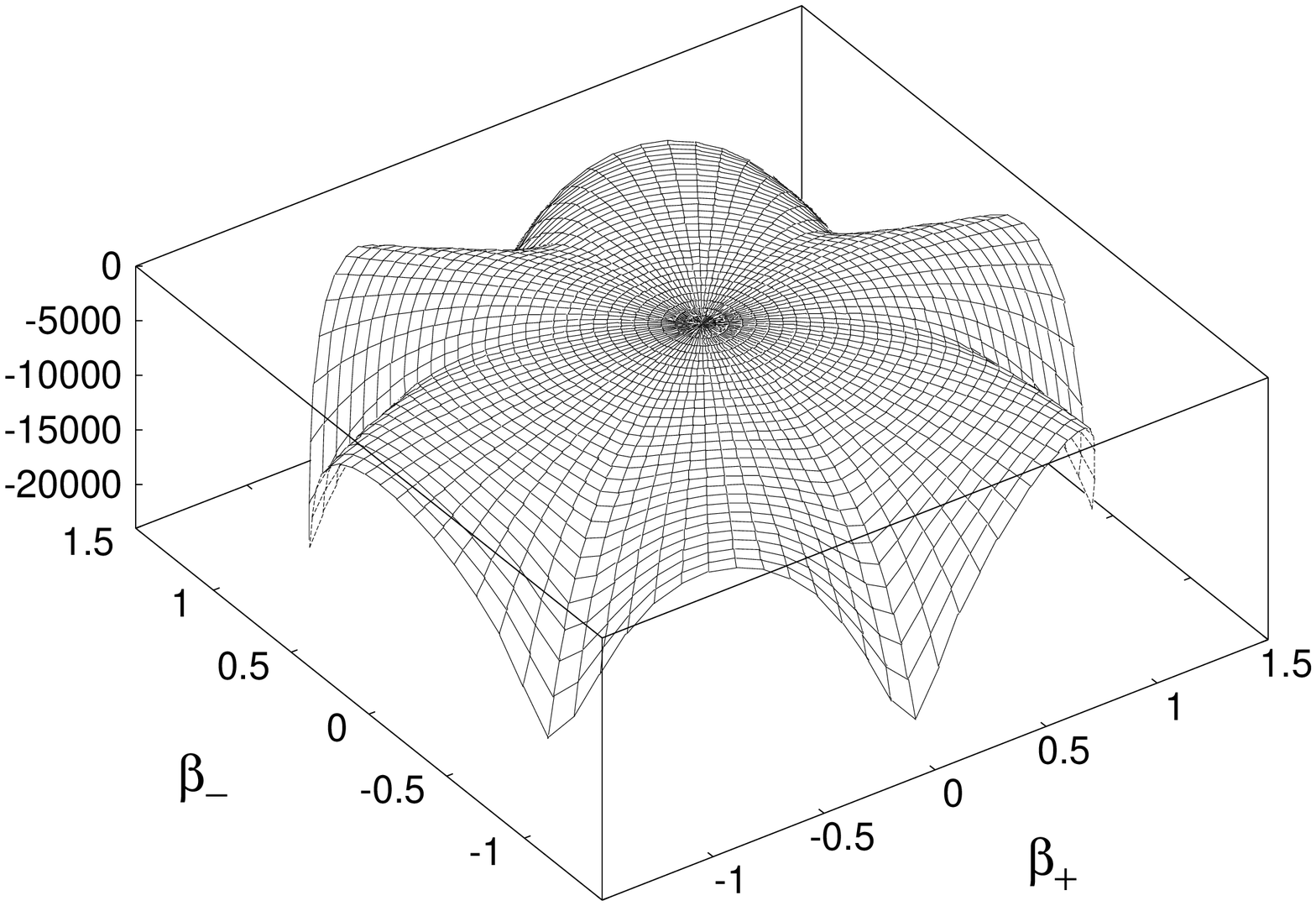}
\end{center}
\caption{The modified effective potential is plotted for $j = 5$ at
volumes 150, 100, 30 in Planck units for $-1.5 \le \beta_{\pm} \le
1.5$.}
\label{ExactFig}
\end{figure}

In Fig.~\ref{ExactFig} are shown three snapshots of the effective
potential at decreasing volumes in the vicinity of the isotropy point
in the anisotropy plane. The potential at larger volumes clearly
exhibits a wall (positive potential) of finite height and finite extent.
As the volume is decreased the wall moves {\em inwards} and its height 
decreases. Progressively the wall disappears completely making the
potential negative everywhere. Eventually, the potential approaches zero 
from below. 

This immediately shows that the classical reflections will stop after
a finite amount of time, rendering the classical argument about chaos
inapplicable. The volume where the transition from the classical
behavior to the modified behavior takes place, i.e.\ the first time
the universe can ``jump over the wall,'' depends on the initial
conditions but it will certainly happen --- the latest at the
elementary discrete volume for the smallest value of the ambiguity
parameter.

Starting at large volumes, the universe will pass through several phases
of Kasner evolution punctuated by reflections. This will continue until
the small volume modification of the effective potential comes into
play. The stability of a Kasner trajectory in the presence of the modified
potential will decide the nature of subsequent deviations after the `last'
Kasner epoch. Pleasingly, the Kasner trajectories turn out to be stable
in the presence of the effective potential \cite{ChaosLQC}. By contrast, 
these are unstable in the presence of the classical potential.

Intuitively, this gives us a natural and consistent picture of the
approach to a classical singularity in a quantum theory with a
discrete structure of space: At larger volume the system follows the
complicated classical evolution with different Kasner phases
interrupted by reflections at the walls. As described before, in the
context of an inhomogeneous space this leads to a fragmentation into
smaller and smaller patches.
Once quantum effects are taken into account, the reflections stop just
when the volume of a given patch is about the size of a Planck volume,
i.e.\ at the scale of the discreteness. Below that scale, a further
fragmentation does not take place and the discrete structure is
preserved.  We again emphasize that the modifications used here become
important at the latest when the Planck scale is reached implying
consistency with the expectations from a discrete structure. Below
this scale, though, the effective classical description should be
superseded by the quantum description. We also emphasize that the
existence of a maximal curvature is a {\em consequence} of quantization 
and {\em not} put in, in an ad-hoc manner. The results thus constitute a 
consistency test for loop quantum gravity.

This also indicates that the results of
\cite{Sing,IsoCosmo,HomCosmo,Spin} which prove a non-singular quantum
evolution of homogeneous models can be generalized to the full theory,
removing also inhomogeneous classical singularities. At the present
stage, however, the results for the general case are to be regarded as
preliminary and have to be supported by more general techniques
directly in inhomogeneous quantum models.

We thank G.~Hossain, R.~Penrose and A.~Rendall for discussions. This
work was supported in part by NSF grant PHY00-90091 and the Eberly
research funds of Penn State.

\end{document}